# New Universal Flavor-Electroweak Physical Constant and Neutrino Type

## E. M. Lipmanov
40 Wallingford Road # 272, Brighton MA 02135, USA

### Abstract

By analogy with the solution of the classical physics problems achieved not by an extension of the established dynamic and symmetry knowledge but by the emergence of a fundamentally new empirical physical constant h leading to quantum mechanics, I address in this paper the known basic problems of lepton mass ratios, neutrino type and electroweak charges in terms of an emerging dimensionless flavor-electroweak physical constant of a new sort. By hints from experimental data, a special value of that new universal constant is suggested $\alpha_o \equiv e^{-5}$. Like the Plank constant h, initially closely related to suggested discrete radiation energy, the new constant $\alpha_o$ is related here to the (known) problem of discrete electric charge. It is observed that the constant $\alpha_o$ determines the mass ratios of Charged Leptons **(CL)**, the mass ratios, absolute mass scale and oscillation mass squared differences of Quasidegenerate **(QD)** neutrinos on the one hand, and the low energy fine structure constant $\alpha \sim \alpha_o$, elementary electric charge $\varepsilon \sim (4\pi\alpha_o)^{1/2}$ and the second electroweak constant $\alpha_W \sim \alpha_o \log \alpha_o^{-1}$, on the other hand. I gathered, organized and commented an interesting system of primary observations made on particle mass and electroweak experimental data. As a result, the new physical constant $\alpha_o$ describes a *fundamental aspect* of low energy phenomenology uniting the electroweak theory with the necessary idea of anthropic selection of the free interaction constant values and flavor freedom mass values.

## 1. <u>Introduction, and two (I and II) guiding suggestions</u>

**1)** The starting phenomenological idea in this work is analogous to the widely discussed in the literature seesaw idea [1]. In the seesaw mechanism, one starts with the unexplained experimental fact of very small neutrino mass and seesaw-relates it to a hypothetical very heavy lepton Majorana mass. In this work, I seesaw-relate the neutrino mass pattern to the other unexplained exceptional experimental fact of the unique lepton mass spectrum — the CL tri-large values of two mass ratios and one mass-ratio hierarchy. The advantage of this approach is that it unites two substantial characteristics of low energy lepton flavor phenomenology: the small mass and unknown neutrino Deviation from Mass-Degeneracy (**DMD**) pattern[1] is seesaw-related to the well known CL masses and their divergent DMD-pattern. The overwhelming advantage of the seesaw mechanism [1] is its close relation to mainstream theoretical trend. The two approaches are not necessarily incompatible[2]. Note that the see-saw [1] is difficult to falsify at attainable energies, and it does not single out a definite neutrino type, though less attention is given to QD-type. In contrast, the approach to small neutrino mass in this work may be definitely falsified, or confirmed (e.g. in cosmology observation analysis), if the neutrino masses are respectively not QD or are QD indeed. Besides, the condition of Majorana neutrinos is crucial in the seesaw mechanism [1] whereas it is favorable, but not crucial, in the neutrino-CL seesaw approach, Appendix A. That approach is supported by the evident analogy between the experimental data neutrino oscillation solar-atmospheric mass-squared-difference hierarchy and CL mass-ratio hierarchy. The unique prediction of QD-neutrinos is important here since if the experiment falsifies it, there is in view no other way to include massive neutrinos in the lepton flavor mass-ratio phenomenology based on the new flavor-electroweak constant $\alpha_o$.

---

[1] The introduced in ref.[2,4] DMD flavor quantities are quantitatively described by $(X_n - 1)$ where $X_n$, $n = 1, 2$, are two lepton mass ratios. Exact mass-degeneracy would be at $(X_n - 1) = 0$.

[2] For a recent discussion, see [18].

**2)** Unlike the one-generation Standard Model, flavor mass physics is still an empirical frontier sector of physics. With that, the empirical forms of the particle flavor mixing matrixes are much more elaborated in the low energy phenomenology (Cabibbo-Kobayashi-Maskawa quark mixing matrix and Pontecorvo-Maki-Nakagawa-Sakata neutrino mixing one) than the particle mass-ratios as the other part of the low energy particle mass matrixes (the third part is the overall mass scale). While the particle mixing matrixes are important for actual calculations, the mass ratios may be more indicative of new flavor physics. There are still several basic problems in low energy lepton flavor mass and neutrino phenomenology, e.g.

(a) What is the pattern of the neutrino masses in comparison with the mass patterns of the CL and quarks?

(b) What is the neutrino mass scale?

(c) Are the main low energy characteristics of the lepton mass spectra in *flavor* physics - the hierarchies of masses $m_n$ and mass-ratios $x_n \equiv (m_{n+1}/m_n)$ (especially the DMD-hierarchies) - connected to known low energy dimensionless quantities of fundamental physics and what may be the connection?

**3)** Among the known elementary particles, the CL and quarks are Dirac particles with divergent hierarchical flavor mass patterns, while the neutrinos are the only particles that may be of Majorana nature and of QD-type. Against this data background, two guiding suggestions are expounded in this paper:

(I) DMD-patterns of the neutrino and CL mass spectra are opposite to each other (DMD-seesaw relation between neutrino and CL mass spectra). This suggestion is rendered quantitatively concrete when the lepton DMD-quantities are expressed in terms of the new constant $\alpha_o \equiv e^{-5}$. The suggestion of QD-neutrinos by relation to CL mass pattern seems artificial only till it is placed against the background of an exactly degenerate pattern with mass ratios $X_n = 1$. In terms of DMD-quantities $(x_n - 1)$, i.e. deviation of CL mass pattern from an exactly mass-degenerate one, that idea looks natural[3].

(II) There is an essential connection between lepton mass hierarchies and low energy electroweak coupling constants[4]. This idea is rendered quantitatively concrete via the new universal intermediate constant $\alpha_o$, as in case of suggestion (I).

Definite answers to the three main lepton flavor physics problems a)-c) are considered below in light of the two guiding suggestions and the basic idea of a new physical constant $\alpha_o$.

In Sec.2, neutrino-CL DMD-seesaw relation is outlined. In Sec.3, a relevant precise connection between the low energy fine structure constant and the CL mass-ratio parameter $\alpha_o$ is discussed and commented as a hint from experimental data on new flavor-electroweak physics. In Sec.4, small QD-neutrino mass scale is estimated from a drastically growing with trend to lower masses lepton mass-ratio hierarchy in terms of $\alpha_o$ without inputs from the oscillation data. In Sec.5, an equation for the neutrino DMD-quantities is obtained from a precise relation (Koide formula) between CL mass ratios; its solution determines

---

[3] Especially in view of the puzzle of tiny neutrino mass.

[4] A relation between electron-muon mass ratio and the low energy fine structure constant is considered in the literature time and again, e.g. [3].

QD-neutrino masses and mass-squared differences. In the extended Sec.6, conclusions are stated and commented. In Appendix A, a generalized Dirac-Majorana DMD-seesaw relation is considered. Appendix B contains a comment on relation between the new constant $\alpha_o$ and the anthropic selection idea. In Appendix C, quark mass ratios are addressed.

## 2. <u>Neutrino-CL DMD-seesaw suggestion</u>

In references [2a] and [2b], exponential lepton (neutrino and CL) mass ratios are inferred from an implied idea of neutrino-CL DMD-oppositeness and the CL experimental mass data in the form

$$X_1 \equiv m_\mu/m_e \cong (2)^{1/2} \exp 5, \quad X_2 \equiv m_\tau/m_\mu \cong (2)^{1/2} \exp(5/2); \qquad (1)$$

$$x_2 \equiv (m_3/m_2)_{nu} = \exp(a\,r), \quad x_1 \equiv (m_2/m_1)_{nu} = \exp(a\,r^2), \quad a\,r \ll 1. \qquad (2)$$

Relations (1) and (2) are *solutions with large and small exponents* respectively (in agreement with the emphasized here neutrino-CL DMD-oppositeness idea) of the generic nonlinear equation (2*) for the CL $(X_n - 1)$ and neutrino $(x_n - 1)$ DMD-quantities:

$$[(X_2 - 1)^2 = A(X_1 - 1)]_{CL}, \quad [(x_2 - 1)^2 = a(x_1 - 1)]_{nu}. \qquad (2*)$$

From experimental data, the CL coefficient in the first equation is equal [4] $A \cong 2^{1/2}(1 - 5e^{-5})$ to within 0.001..

There is a condition [2b] on the coefficient $a > a_{min} = 0.85$ in the neutrino equation (2*) leading to the important physical restriction on the neutrino parameter $r \ll 1$.

The three neutrino mass eigenvalues are $m_1 < m_2 < m_3$.

It follows from (2) that the neutrino masses are of QD-type and the parameter $r$ in the neutrino mass ratios has a double physical meaning of: 1) neutrino DMD-ratio (hierarchy), and 2) solar-atmospheric neutrino oscillation hierarchy parameter:

1) $r \cong (x_1^2-1)/(x_2^2-1)$, 2) $r \cong (\Delta m^2_{sol}/\Delta m^2_{atm})$. (3)

The distinct inferences (3) are independent of the numerical value of the exponential coefficient 'a' in the neutrino mass ratios (2).

With mass ratios (2), the absolute value of QD-neutrino mass scale is given by

$m_\nu \cong (\Delta m^2_{atm}/2ar)^{1/2} \cong (\Delta m^2_{sol}/2ar^2)^{1/2} \cong \Delta m^2_{atm}/(2a\Delta m^2_{sol})^{1/2}$, (4)

It contains one not fully determined above coefficient[5] 'a'.

It is argued in refs.[2] and [4] that the small exponential factor $r$ in the QD-neutrino mass ratios (2) should be related to the large CL exponential factor in (1):

$r \cong 5\exp(-5) \cong 0.034 \ll 1$. (5)

Relation (5) is a special quantitative concretization of the guiding neutrino-CL DMD-seesaw idea. Estimation (5) is in good agreement with the estimations of the parameter $r$ in refs. [5] and [6] from oscillation experimental data.

With the definition of a new universal constant $\alpha_o$

$\alpha_o \equiv e^{-5}$, (6)

relations (1) and (5) are given by

$X_1 \cong (2)^{1/2}/\alpha_o$, $X_2 \cong (2/\alpha_o)^{1/2}$, (1')

$r \cong 5\alpha_o = \alpha_o \log\alpha_o^{-1}$. (5)

Since the CL mass ratios are large and the QD-neutrino ones are close to unity, the DMD-quantities of the CL and neutrinos are respectively large and small

$(X_1^2-1) \cong 2/\alpha_o^2$, $(X_2^2-1) \cong (2/\alpha_o)$, (7)

---

[5] With the relations $m_\nu^2 \cong 7\Delta m^2_{atm}$ (see Sec.5), (4), (5) and oscillation mass squared differences from [17], it follows $(ar) \cong 1/14$, $a \cong 2$. And so, from the third relation (4): $m_\nu \cong (0.11 \div 0.18)eV$. It implies that the initial generic approximate equation for the lepton (CL and neutrino) DMD-quantities is $(x_2^2-1) \cong 2(x_1^2-1)$, instead of $(x_2-1) \cong (2)^{1/2}(x_1-1)$ as in (2*) with a = (2)$^{1/2}$, i.e. the primary DMD-quantities should be defined through 'mass-squared-ratios' $x_n^2$ instead of $x_n$, comp. hep-ph/0304207 and [2b].

$$(x_2{}^2-1) \cong 2a\,r, \quad (x_1{}^2-1) \cong 2a\,r^2, \qquad\qquad (8)$$

in agreement with the lepton DMD-seesaw idea. That idea explains [2, 4] the *remarkable experimental fact* of analogously large[6] hierarchies of neutrino oscillation mass-squared differences

$$r \equiv (x_1{}^2-1)/(x_2{}^2-1) \cong (\Delta m^2_{sol}/\Delta m^2_{atm}) \cong 5\,\alpha_o \equiv \alpha_{oW} \qquad (9)$$

and CL mass-squared ratios ,

$$R \equiv (X_2{}^2-1)/(X_1{}^2-1) \cong (m_\tau/m_\mu)^2/(m_\mu/m_e)^2 \cong \alpha_o\,. \qquad (10)$$

The analogy between relations (9) and (10) is interesting: in both cases, lepton DMD-hierarchy parameters (left sides) are close to the parameters $\alpha_{oW}$ and $\alpha_o$ which in turn are close to the experimental values of low energy (pole[7]) electroweak gauge coupling constants $\alpha_W \approx \alpha_{oW}$ and $\alpha \approx \alpha_o$ respectively ($\alpha$ is the fine structure constant, $\alpha_W$ - its semiweak analogue, see bellow).

### 3. <u>The constant $\alpha_o \equiv \exp(-5)$ and low energy fine structure constant</u>

<u>**1.**</u> Why the emergence of the exponential $e^{\pm 5}$ in the CL mass ratios? I argue in favor of an essential answer - the exponential $e^{\pm 5}$ is related to new flavor-electroweak physics and is a new universal dimensionless physical constant [4]. Let us start with a very approximate relation for the fine structure constant and trace the way to the precise one,

1) The relation

$$\alpha \cong \alpha_o \equiv e^{-5} \cong 1/148.4, \qquad\qquad (11)$$

is correct to within ~8% ($\alpha \cong 1/137$). Approximation (11) is prompted by the lepton mass-ratios above, relation (9) $\alpha_W \approx$

---

[6] Both large and small values of DMD-hierarchy describe 'large' physical hierarchy, unlike the case of DMD-quantities themselves.

[7] See Eq.(43) below.

5 exp(-5) for the semiweak constant, empirical value [8] of the Weinberg mixing angle $\sin^2\theta_W \cong 0.2$ and electroweak theory [9] connection $\alpha = \alpha_W \sin^2\theta_W$. The approximation (11) is a remarkable one, with it and (5)-(10), we get

$$\alpha \approx \alpha_o, \ \sin^2\theta_W \approx 1/\log\alpha_o^{-1} = 0.2, \ \alpha_W \approx \alpha_{oW} \equiv \alpha_o \log\alpha_o^{-1}, \qquad (12)$$

$$m_\mu/m_e \cong 2^{1/2}/\alpha_o, \ m_\tau/m_\mu \cong (2/\alpha_o)^{1/2}, \ R \cong \lambda'\alpha_o, \qquad (13)$$

$$(m_3/m_2-1)_{nu} \cong a\,r, \ (m_2/m_1-1)_{nu} \cong a\,r^2, \ r = \lambda\,\alpha_{oW}, \qquad (14)$$

$$\alpha_{oW} \equiv 5e^{-5}, \ \lambda \approx 1, \ \lambda' \approx 1,$$

where $r$ and R are the DMD-ratios (hierarchies) of respectively neutrinos and CL. Relations (12)-(14) are correct to within ~(1-8)%, but they seem meaningful and suggestive, as the starting relation (11) does. The five independent dimensionless coupling constants of the low energy electroweak lepton interactions, with the electromagnetic field and $W^\pm$-field (gauge interactions) and scalar Higgs-field interactions ($f_e$, $f_\mu$ and $f_\tau$ — coupling constants with the scalar field of the electron, muon and tauon respectively)[8],

$$f_\mu \cong (2^{1/2}/\alpha_o)\,f_e, \ f_\tau \cong (2/\alpha_o)^{1/2}f_\mu, \qquad (15)$$

are united via the parameter $\alpha_o$. One coupling constant with the scalar field is left free (e.g. $f_e$, i.e. electron mass $m_e$).

2) Observe now with $(\alpha^{-1})_{Data} \cong 137.036$ from [8] that the following extension of relation (11)

$$\exp 2\alpha\,\log(\exp\alpha/\alpha) \cong \log(1/\alpha_o) \qquad (16)$$

is accurate to within ~$3 \times 10^{-6}$ as an experimental fact, and it is suggestive: i) Considering (16) as an equation for the unknown $\alpha$, the solution is

$$\alpha \cong 1/137.0383, \quad (\alpha - \alpha_{Data})/\alpha_{Data} \cong -1.7 \times 10^{-5}, \qquad (17)$$

compared to $[\alpha_o - \alpha_{Data}]/\alpha_{Data} \cong -0.08$.

---

[8] For the CL: $m_\ell = f_\ell \langle\varphi\rangle$, $\ell = e, \mu, \tau$, $\langle\varphi\rangle$ is the vacuum expectation value of the scalar field $\varphi$.

ii) A definite and interesting indication from Eq.(16) is that, after exponentiation

$$(\exp\alpha /\alpha)^{\exp 2\alpha} = 1/\alpha_o, \qquad (16')$$

the difference between the right and left sides in (16') is equal to $(\alpha/\pi)$ to within $\sim 2 \times 10^{-3}$, a significant hint from the experimental data.

3) Taking this hint, we get a further extended relation

$$\exp 2\alpha \, \log(\exp\alpha /\alpha) \cong \log(1/\alpha_o - \alpha/\pi) \qquad (18)$$

between the source-value $\alpha_o$ and the highly accurate data value of the fine structure constant [8]:

$$(1/\alpha)_{Data} = 137.03599911(46). \qquad (19)$$

With (19), relation (18) is accurate to within $\sim 6 \times 10^{-9}$.

Considering (18) as an equation for the unknown $\alpha$, the solution is given by

$$\alpha \cong 1/137.0359948, \quad (\alpha - \alpha_{Data})/\alpha_{Data} \cong 3.1 \times 10^{-8}. \qquad (20)$$

It differs from the central data value of the fine structure constant (19) only by about 10 S.D.

Note that the additional term $\alpha/\pi$ on the right side of (18) is like a "perturbation term" added to the "nonperturbative" basic relation (16').

4) Finally, a possible second "perturbative" term (e.g. $\alpha^2/4\pi$) raises the accuracy of relation (18) by about one order of magnitude, and transforms it into a highly accurate nonlinear equation for the unknown $\alpha$:

$$\exp 2\alpha \, \log(\exp\alpha /\alpha) = \log(1/\alpha_o - \alpha/\pi + \alpha^2/4\pi), \qquad (21)$$

or, comp. (16'),

$$(\exp\alpha /\alpha)^{\exp 2\alpha} + \alpha/\pi - \alpha^2/4\pi = \exp 5 \equiv 1/\alpha_o. \qquad (21')$$

Indeed, the solution of Eq.(21) is given by

$$\alpha \cong 1/137.03599901, \quad (\alpha - \alpha_{Data})/\alpha_{Data} \cong 0.7 \times 10^{-9}. \qquad (22)$$

It agrees with the central data value of the fine structure constant at zero momentum transfer $\alpha_{Data}$ (19), to within ~0.2 S.D.

In summery, equation (21) is an accurate equation for the fine structure constant $\alpha$ at zero momentum transfer with the exponential exp5 on the right side of Eq.(21') as the source of the precise numerical solution (22). A more compact form of Eq.(21') is given by

$$(\exp\alpha /\alpha)^{\exp 2\alpha} + (\alpha/\pi)\exp(-\alpha/4) = \alpha_o^{-1}. \qquad (21'')$$

Its solution is $\alpha \cong 1/137.03599900$.

**2.** Based on the new measurement of the (g-2)-factor in the experiment [19], the recently discovered [20] precision experimental value of the fine structure constant[9],

$$(1/\alpha)_{exp} = 137.035999710(96), \qquad (19')$$

gets a perfect fit by the solution of the nonlinear equation

$$(\exp\alpha /\alpha)^{\exp 2\alpha} + (\alpha/\pi) - (\alpha/\pi)(\alpha_o/\pi) = 1/\alpha_o = \exp 5. \qquad (21*)$$

In contrast to Eq.(21'), equation (21*) is characterized by two special terms - one main exponential nonlinear in $\alpha$ term and a much smaller linear in $(\alpha/\pi)$ "perturbative" term. The accurate numerical solution of Eq.(21*) is given by

$$\alpha \cong 1/137.0359997426. \qquad (22*)$$

A more compact form of Eq.(21*) is

$$(\exp\alpha /\alpha)^{\exp 2\alpha} + (\alpha/\pi)\exp(-\alpha_o/\pi) = 1/\alpha_o \qquad (21**)$$

---

[9] Another estimation of the fine structure constant in ref.[21], with data inputs from the same experiment [19], is given by $\alpha_{exp}$= 1/137.035999709(96); it is in exact agreement with the result (19') from [20]. I would like to thank M. Passera for interest and the information.

with the almost equally accurate solution

$$1/\alpha = 137.0359997372. \qquad (22**)$$

As a result, the solutions (22*) and (22**) agree with the most accurate to date experimental value (19') of the fine structure constant from ref.[20] to within a remarkable accuracy ~+2x10$^{-10}$, or ~+0.3 S.D.

It may be curious to comment on the origin of the accurate equation (21*) for the fine structure constant $\alpha$. In electroweak theory the constant $\alpha \equiv \alpha(Q^2 = 0)$ is a free parameter and is *determined only by the experimental value* $\alpha_{exp}$. This status of the fine structure constant $\alpha$ is not changed by the Eq.(21*), but $\alpha$ is here determined only by the new universal constant $\alpha_o$ instead of $\alpha_{exp}$. Since the constant $\alpha$ appears close to $\alpha_o$ (~8%), see (11), the main term of the connection between $\alpha$ and $\alpha_o$ follows from the assumption that it is a *nonlinear light dressing* of $\alpha$ by the natural here exponential factors $(\exp \alpha)^n$, $n = 0,1,2…$ Then, a few steps of selection (especially simple in logarithmic form, see (16)) lead to the nonlinear Eq.(16'), and with hints from new data [20] for the linear term – to Eq.(21*).

**3.** The interesting old question of where does the specific numerical value (dimensionless number) of the fine structure constant at zero momentum transfer come from[10] may have a new answer. With the relations above, a unique connection between the fine structure constant $\alpha$ and the CL flavor parameter $\alpha_o$ does exist, at least to within a few S.D.  If the true numerical value of the fine structure constant at $Q^2 = 0$ is reduced to integer 5 (alpha-genesis[11]) through the source value $\alpha_o \equiv e^{-5}$, all gauge coupling constants of the Standard Model might be expressed through the parameter $\alpha_o$ by renormalization group equations and Grand Unification [10].

---

[10] Fundamental physical meaning is attached to low energy phenomenology in our universe placed against a multiverse background by the anthropic principle in ref.[11].

**4.** We should point out that there is also another reason why the relations (12)-(14) are particularly interesting: they afford a special answer to the obvious question of why the muon and tauon are needed in the low energy physics region $E < m_e/\alpha^2$. With $\alpha_o$ as a universal constant in flavor physics and finite electron mass $m_e$, the $m_\mu$ and $m_\tau$ would be extremely large and the neutrino mass would be $m_\nu = 0$ in the limit $\alpha = \alpha_o \rightarrow 0$, but the electroweak force would disappear. So, with the universal parameter $\alpha_o$ in the flavor phenomenology above the muon and tauon with finite masses are necessary for the electron-flavor generation particles to have bound states.

**5.** The parameter $\alpha_o$ has here a dual functional meaning: from (21) it is the source of the fine structure constant magnitude in QED, and at the same time from (10) it is the source of the CL mass-ratio hierarchy R in flavor physics, $R \cong 0.98\,\alpha_o$. The local gauge symmetry of QED determines conservation of electric charge and the interaction of the electron with electromagnetic field, but not the value of the elementary electric charge $\varepsilon = (4\pi\alpha)^{1/2}$.

---

[11] As a finite number, the value $\alpha_o$ uniquely determines the empirical value $\alpha_{Data}$ and, with the QED renormalization group equations, all the further growing values of the fine structure constant at higher momentum transfers.

The minimal observable value of the fine structure constant $\alpha$ at the particular momentum transfer $Q^2=0$ is a very special physical quantity because of its role in nonrelativistic quantum mechanics of bound states enabling Life and Consciousness. Its unique relation to the exponential $\alpha_o \equiv e^{-5}$ is a possible solution to the anthropic principle problem for the fine structure constant in our universe.

That remaining problem likely needs entirely new physics. In this work, the absolute value ε (in natural units) is generated by the new universal physical constant $\alpha_o$.

**6.** In summery, the fine structure constant value and the absolute value of the elementary electric charge are encoded[12] in the ripples (hierarchies) on an imaginary exactly mass-degenerate pattern of CL flavor copies as a background on which the CL mass-flavor physics is displayed. In the present phenomenology, the elementary electric charge value and the QED interactions of CL would disappear without CL copies and their mass pattern ripples.

## 4. <u>QD-Majorana-neutrino mass scale from drastic lepton mass-ratio hierarchy</u>

In the QD-neutrino scenario, the lepton mass spectrum contains 4 mass-degenerate Majorana mass levels: three two-fold exactly-mass-degenerate[13] CL mass levels $m_\tau$, $m_\mu$ and $m_e$ (three carrying charge Dirac states) plus one three-fold Quasi-Degenerate Majorana-neutrino mass level $m_\nu$. The three CL mass levels are highly hierarchical, with the hierarchy-rule approximately described by Eq.(13) in terms of powers of the universal parameter $\alpha_o$. Can this CL mass-ratio hierarchy be extended so to include the fourth very low QD-neutrino mass level $m_\nu$? An affirmative answer, ref. [4], is possible only in

$$X_1 \cong 2^{1/2}(\alpha_o^{-1}), \ X_2 \cong (2\alpha_o^{-1})^{1/2}, \ R \cong \alpha_o, \ \alpha_o^{-1} = f(\alpha),$$

with the exact explicit expression (21'') for the function $f(\alpha)$.

---

[12] In accordance with the solution (1) for CL mass ratios, and (10) for the DMD-hierarchy parameter R, these ratios can be rewritten in the form

[13] This exact mass-degeneracy (symmetry) cannot be broken so far as there are no interactions that violate electric charge conservation, in contrast to conservation of lepton charge.

the QD-neutrino scenario - it is a factorial hierarchy, which extends the sequence in (13):

$$m^2_{\ell+1}/m^2_\ell \cong \alpha_o^{\ell!}/2. \qquad (23)$$

The notations are $m_1 = m_\tau$, $m_2 = m_\mu$, $m_3 = m_e$ and $m_4 = m_\nu$. Inclusion of the mass level $m_\nu$ into the CL mass-ratio hierarchy is enabled by the complementary remote view at the QD-neutrino mass pattern as an exactly mass-degenerate one. With very small finite neutrino mass, the hierarchy (23) fits well the empirical picture of the lepton mass pattern on the whole.

The factorial mass-ratio hierarchy between four lepton mass levels (23) is represented by three terms

$$m^2_\mu/m^2_\tau \cong \alpha_o/2, \; m^2_e/m^2_\mu \cong \alpha_o^2/2, \; m^2_\nu/m^2_e \cong \alpha_o^6/2. \qquad (24)$$

Hence, the absolute value of the QD-neutrino mass scale is given by

$$m_\nu \cong \alpha_o^3 \, m_e/2^{1/2} \cong 0.11 \text{ eV}. \qquad (25)$$

Because of the high power of the constant $\alpha_o$ in (25), a change of the value $\alpha_o$ to the exact value of the low energy fine structure constant $\alpha$ may lead to a noticeable increase of the neutrino mass scale

$$m_\nu \cong \alpha^3 \, m_e/(2)^{1/2} \cong 0.14 \text{ eV}. \qquad (26)$$

The estimations (25) and (26) are compatible with the cosmological neutrino bounds [12] for QD-neutrinos, $m_\nu < 0.14$ eV at 95% C.L.

## 5. **Neutrino DMD-quantities from Koide formula**

An interesting accurate low energy relation between physical CL mass ratios is the Koide formula [13]:

$$(m_e + m_\mu + m_\tau) = 2/3[(m_e)^{1/2} + (m_\mu)^{1/2} + (m_\tau)^{1/2}]^2. \qquad (27)$$

With the notations for the CL mass ratios $X_1 = m_\mu/m_e$, $X_2 = m_\tau/m_\mu$, rewrite (27) in the form

$$(1/X_1 + 1 + X_2)\,3/2 \;=\; [1/(X_1)^{1/2} + 1 + (X_2)^{1/2}]^2. \qquad (28)$$

This is an accurate empirical CL three-flavor mass-ratio equation. In accordance with neutrino-CL DMD-seesaw idea, the mass scale $m_\nu$ of QD-neutrinos may be inferred from the relation (28) by the substitution [4]:

$$X_{1,2} \to (x_{2,1} - 1) \cong (\Delta m^2{}_{atm};\ \Delta m^2{}_{sol})/2m_\nu{}^2 \qquad (29)$$

for $n = 1,2$ respectively. An equation for QD-neutrino DMD-quantities $(x_n - 1)$ is deduced from (28) and (29) in the form:

$$[(x_2-1)^{-1} + 1 + (x_1-1)]\,3/2 = [(x_2-1)^{-1/2} + 1 + (x_1-1)^{1/2}]^2. \qquad (30)$$

With the condition $(x_1-1) \ll 1$, this equation is reduced to

$$[1/(x_2-1) + 1]\,3/2 \cong [1/(x_2-1)^{1/2} + 1]^2. \qquad (30')$$

The important main result for the larger DMD physical quantity $(x_2-1)$ of the neutrino mass pattern[14] follows from Eq.(30'), the neutrino mass-ratios (2) and absolute mass scale (4):

$$(x_2-1) \cong 1/14: \quad (ar) \cong 1/14, \quad m_\nu{}^2 \cong 7\,\Delta m^2{}_{atm}. \qquad (31)$$

The same result is obtained from the initial Eq.(27)-(28) by discarding electron mass and using the substitution

$$X_{1,2} \to (x_{1,2} - 1) \qquad (29')$$

instead of the one in (29).

With 99% C.L. ranges [17] of neutrino oscillation data mass-squared differences

$$\Delta m^2{}_{atm} = (2.1 \div 3.1)\times 10^{-3}\ eV^2, \quad \Delta m^2{}_{sol} = (7.2 \div 8.9)\times 10^{-5}\ eV^2 \qquad (32)$$

and (31), the QD-neutrino mass scale is given by

$$m_\nu \cong (0.12 \div 0.15)\ eV, \qquad (33)$$

with best fit value

$$(m_\nu)_{bf} \cong 0.13\ eV. \qquad (34)$$

---

[14] The Koide formula with substitutions (29) serves here the main purpose to determine the numerical value of the larger DMD-quantity $(ar)$ in the structures (2). It is one of the two special physical quantities of the QD-neutrino mass pattern, with $(ar^2)$ - the other one.

Note that these estimations of QD-neutrino mass scale follow not directly from the Koide mass-ratio formula (28), but from the modified by substitution (29) new equation (30) for the QD-neutrino DMD-$(x_n-1)$-quantities.

By combining the relation between neutrino mass $m_\nu^2$ and atmospheric mass-squared difference $\Delta m^2_{atm}$ (31) with the entirely independent estimation of $m_\nu$ from Sec.4, (25) and (26), an interesting quantitative prediction for $\Delta m^2_{atm}$ results,

$$\Delta m^2_{atm} \cong (\alpha_o^6 \div \alpha^6)\, m_e^2/14 \cong (1.8 \div 2.8) \times 10^{-3}\ eV^2, \qquad (35)$$

in surprisingly good agreement with the 99% CL neutrino oscillation data (32). With $r \cong \alpha_{oW}$, the solar mass-squared difference is given by

$$\Delta m^2_{sol} \cong 0.034\, \Delta m^2_{atm} \cong (5.9 \div 9.5) \times 10^{-5}\ eV^2, \qquad (36)$$

also in good agreement with neutrino oscillation data (32).

It should be noted that no input information from neutrino oscillation data is used in the estimations (35) and (36) - only inferences from neutrino-CL DMD-seesaw relation and the CL mass data relation (Koide formula).

The estimation of the neutrino mass scale in Sec.4, Eqs.(25) and (26), is determined by lepton mass-ratio hierarchy (23) and is independent of the neutrino oscillation data; the estimation of $m_\nu$ in (33) is determined mostly by the atmospheric neutrino oscillation mass-squared difference $\Delta m^2_{atm}$. Remarkably, both estimations of the QD-neutrino mass scale are in fair agreement with each other.

Notice: The empirical equations (2*) and (28) for CL mass ratios $X_1$ and $X_2$ are a compatible system of two equations for two unknowns. A solution of this system for CL mass ratios in terms of the universal parameter $\alpha_o$ is given by

$$X_n \cong 2^{1/2}(1-\alpha_{oW})\,[(\exp 3\alpha_o)/\alpha_o]^{1/n}, \quad \alpha_{oW} = 5\alpha_o, \ n=1,2. \qquad (1'')$$

From the experimental data, the accuracy of the muon-electron mass-ratio $X_1$ from (1'') is $\sim 9 \times 10^{-4}$, while the accuracy of the tauon-muon mass-ratio $X_2$ is $\sim 5 \times 10^{-5}$. The interesting feature of these bare CL mass-ratios is that they satisfy the Koide equation (28) to within a remarkable accuracy $\sim 5 \times 10^{-6}$ despite the evident much smaller accuracy of both CL mass ratios themselves.

## 6. <u>Conclusions and comments</u>

Experimental and phenomenological indications of a new primary flavor-electroweak physical constant $\alpha_o \equiv e^{-5}$ are observed. With that constant, the functional dependence on integer 5 of the DMD-hierarchies **$r$ and $R$** of the QD-neutrinos and CL approximately[15] resembles that of the low energy electroweak coupling constants **$\alpha_W$ and $\alpha$.**

In the electroweak theory of leptons [9] there is no connection between the lepton masses and electroweak charges.

Though there are no mass-charge relations for individual particles in the present phenomenology, but a kind of 'nonlocal' connections are observed between the DMD-hierarchies R and *$r$* as flavor (family) mass quantities of CL and neutrinos, on the one hand, and the two electroweak charges, on the other hand.

Conclusions, supported by considered flavor and electroweak experimental data, are commented below.

*<u>1)</u> New universal flavor-electroweak constant $\alpha_o \equiv e^{-5}$.*
The dimensionless constant $\alpha_o$ is indicated by parametrizing the accurate experimental values [8] of CL mass ratios $X_1 = (m_\mu / m_e)$,

---

[15] "…in the description of nature, one has to tolerate approximations, and that even work with approximations can be interesting and can sometimes be beautiful" - P. A. M. Dirac, Scientific autobiography, in *History of 20th Century Physics*, NY (1977).

$X_2{}^2 = (m_\tau / m_\mu)^2$ and pole electroweak constants $\alpha(Q^2 = 0)$ and $\alpha_W(Q^2 = M_W{}^2)$ in an universal form: $A_i \exp(\pm N_i)$. From these data, the integer 5 is a definite common dominating part of the exponents $N_i = 5(1 + \Delta_i)$ with $|\Delta_i| \ll 1$, while the coefficients $A_i$ are of order 1, see table:

| Experiment. values $\rightarrow$ | $X_1$ $\cong 206.768$ | $X_2{}^2$ $\cong 282.85$ | $\alpha^{-1}(Q^2=0)$ $\cong 137.036$ | $\alpha_W{}^{-1}(Q^2=M_Z{}^2)$ $\cong 29.79$ |
|---|---|---|---|---|
| $A_i \equiv$ | $\sqrt{2}$ | 2 | 1 | $1/5 \equiv 1/\log \alpha_o{}^{-1}$ |
| $(\Delta_i)_{\exp} \cong$ | −0.003 | −0.01 | −0.016 | 0.0007 |

In Sec.3, the precise experimental value [19] of the fine structure constant at $Q^2=0$ is quantitatively derived as solution (22) at the high ppb accuracy level in terms of the universal constant $\alpha_o$.

The introduced new physical constant $\alpha_o$ serves two goals:
i) Unites flavor physics data values with electroweak ones,
ii) Unites EW theory with the anthropic idea for selection of the free values of coupling constants and particle mass ratios.

All considered in this paper dimensionless quantities — charged lepton mass ratios, QD-neutrino mass ratios, lepton DMD-quantities and DMD-hierarchies, low energy electroweak coupling constants $\alpha$ and $\alpha_W$, and the three CL Yukawa coupling constants[16], plus an exact equation for $\alpha$ — are expressed through the one parameter $\alpha_\circ$, e.g. (7), (8), (12)-(14), (21*), (24), (35), (36) and (37)-(43) below. It is a large system (comprising all known dimensionless basic lepton mass and EW interaction constants, without gravity) of experimental indications pointing to the suggestion of a new flavor-electroweak physical constant $\alpha_\circ$ and relating the three-flavor physics to the well established one-generation electroweak physics. There is no incongruity of this basic suggestion with the highly successful Standard Model of elementary particles, which does not describe the experimental data of flavor physics.

---

[16] Note also two low energy approximate empirical relations
$$m_\tau \cong \alpha \langle\varphi\rangle, \qquad m_W{}^2 \cong \pi\alpha_{oW} \langle\varphi\rangle^2,$$
Here $\langle\varphi\rangle \cong 246$ GeV is the vacuum expectation value of the Higgs scalar field. These relations plus (24) determine all CL Yukava coupling constants, and the W-boson mass $m_W$, through the parameter $\alpha_o$.

**_2)_** _Essential connections between the lepton DMD-hierarchies and low energy electroweak charges_ - one of the two guiding ideas in this paper. It is rendered quantitatively concrete in terms of the new constant $\alpha_\circ$, and generates answers to many questions concerned with $\alpha_\circ$—related values of the lepton flavor and neutrino oscillation experimental data, e.g.:

(1) Why extra CL copies beyond the electron are needed in the low energy region $E < m_e/\alpha_\circ^2$ ?

(2) Why CL mass ratios, and especially their hierarchy R, are mainly expressed through the parameter $\alpha_\circ$, and why does the same experimental parameter precisely determine the fine structure constant $\alpha$ at $Q^2 = 0$ ?

(3) Why the small value of the solar-atmospheric hierarchy parameter _r_ may be close to the low energy weak coupling constant squared $\alpha_W$ ?

(4) Why does the hierarchy of neutrino oscillation mass-squared differences $[\Delta m^2_{sol}/\Delta m^2_{atm}]$ resemble the known from the experimental data CL mass-ratio hierarchy $[(m_\tau/m_\mu)^2/(m_\mu/m_e)^2]$? Both are large[17] hierarchies $(r^{-1}, R^{-1})_{exp} \gg 1$.

(5) Why both the fine structure constant $\alpha$ and the semiweak constant $\alpha_W$ are determined by the parameter $\alpha_o$ at analogous (pole) values $Q^2$ of momentum transfers?

That idea is described in Secs.3 and 2 by Eq.(21), connecting $\alpha_o$ with $\alpha$, and two pairs of conformable (analogous) hierarchical lepton DMD-structures:

$$(X_2^2-1)_{\mathbf{CL}} \cong 2\,(\alpha_o^{-1}), \quad (X_1^2-1)_{\mathbf{CL}} \cong 2\,(\alpha_o^{-1})^{\mathbf{2}},$$

$$(x_2^2-1)_{\mathbf{nu}} \approx 4\,(\alpha_{oW}), \quad (x_1^2-1)_{\mathbf{nu}} \approx 4\,(\alpha_{oW})^{\mathbf{2}}, \quad \alpha_{oW} \equiv 5\alpha_o. \qquad (37)$$

Note that there are no small CL DMD-quantities and no large QD-neutrino ones.

The lepton masses and EW charges are independent empirical parameters in the one-generation EW theory [9]. Here, these charges are encoded in the lepton DMD-quantities and so are *connected with the lepton families as physical systems*. In case of one particle generation the effect of the new constant $\alpha_o$ would be negligible. As shown, its real emergence as a physical constant is in case of three generations.

*3)* *Neutrino oscillation solar-atmospheric hierarchy parameter* $\Delta m^2_{sol}/\Delta m^2_{atm}$ *and the low energy semiweak coupling constant* $\alpha_{\mathbf{W}}$. It is argued in [2] and [4] and Sec.2, especially Eq.(5), that the magnitude of the neutrino oscillation solar-atmospheric hierarchy parameter should be close to the constant[18] $\alpha_{\mathbf{oW}} \equiv 5\,\alpha_{\mathbf{o}}$,

---

[17] Unlike the DMD-quantities themselves, large and small values of the *ratios* of DMD-quantities describe the same 'large DMD-hierarchies' (a generic characteristic of lepton, CL and neutrino, mass patterns) in contrast to 'order 1 DMD-hierarchies' in case of near geometrical mass patterns (a probable characteristic of flavor quark mass patterns [14]).

$$r = \lambda \, \alpha_{oW} \cong 0.034 \, \lambda, \quad \lambda \cong 1. \tag{38}$$

In addition, this inference is supported by the relevant parallelism

$$R = \lambda' \alpha_o, \quad r = \lambda \, \alpha_{oW}, \quad (\lambda, \lambda') \approx 1, \tag{39}$$

where the second relation is a discussed above hypothesis, while the first one is a fact $\lambda' \cong 0.98$. The analogy (parallelism) in (39) is possible in case of QD-neutrinos[19] and three flavors[20] where the parameter $r$ has a double physical meaning of 1) neutrino DMD-hierarchy, and 2) solar-atmospheric oscillation hierarchy parameter: $r = (x_1^2 - 1)/(x_2^2 - 1) \cong (\Delta m^2_{sol}/\Delta m^2_{atm})$.

Mention two other possibilities:

1) With the SM prediction [15] $\sin^2\theta_W|_{Q2=0} \cong 0.2383$:

$$r = \lambda_1 \, \alpha_W|_{Q2=0} \cong 0.031 \, \lambda_1, \quad \lambda_1 \approx 1. \tag{40}$$

2) By shifting the value $\alpha_o$ to $\alpha = \alpha_{Data}$ in (12):

$$r \cong \lambda_2 \, [\alpha \log(1/\alpha)]_{Q2=0} \cong 0.036 \, \lambda_2, \quad \lambda_2 \approx 1. \tag{41}$$

These estimations are examples of different concrete interpretations of the suggested connection between the neutrino DMD-hierarchy $r$ and the dimensionless weak coupling constant $\alpha_W$.

---

[18] The constant $\alpha_{oW}$ seems analogous to the source value $\alpha_o$ of the fine structure constant $\alpha$. The difference is that $\alpha_o < \alpha(Q^2=0)$, while $\alpha_{oW} > \alpha_W(Q^2=0)$. Since the running coupling constant $\alpha_W(Q^2)$ is increasing in the space between $Q^2=0$ and $Q^2 \approx M_W^2$, the condition $\alpha_W(Q_1^2) = \alpha_{oW}$ may be realized at some value $Q_1^2 > 0$ which in fact is $Q_1^2 \approx M_W^2$.

[19] Also in case of inverted hierarchy where two neutrino mass levels are almost degenerate, it is not considered here.

[20] If the number of flavors is more than three, there would be more than two DMD-hierarchies while there are only two independent electroweak gauge coupling constant $\alpha$ and $\alpha_W$. So, the fact of three flavors is very natural in the outlined phenomenology.

With $\lambda = \lambda_{1,2} = 1$, the estimations (38) and (41) are in better agreement with the data values, see [11-13].

   **4)** *Special analogy between the electroweak constants $\alpha$ and $\alpha_W$ and the universal constant $\alpha_o$:*

$$\alpha(Q^2 = 0) = f^{-1}(\alpha_o^{-1}), \quad \alpha_W(Q^2 \cong M_W^2) \cong \alpha_o \log(\alpha_o^{-1}) \equiv \alpha_{oW} \qquad (42)$$

$(\alpha_o^{-1} = f(\alpha)$ is an explicit function from Eq.(21*)). The momentum transfers in (42) are equal to the pole values of the photon and W-boson propagators $Q^2 = 0$ and $Q^2 = M_W^2$, respectively. That pair of analogous connections between the basic electroweak constants $\alpha$ and $\alpha_W$ and the emerging dimensionless constant $\alpha_o$ in (42) is in turn analogous to the other pair of analogous relations (39) between the lepton DMD-hierarchies ($r$ and R) and that constant $\alpha_o$. This double pair analogy,

$$R \cong \alpha_o \cong \alpha, \quad r \cong \alpha_{oW} \cong \alpha_W, \qquad (43)$$

is in support of the definition $\alpha_o$ as a universal physical constant.

   **5)** *The muon and tauon flavor counterparts of the electron are unavoidable.*

With the universal parameter $\alpha_o$ in the framework of low energy flavor phenomenology, a physically meaningful limiting case with only one flavor generation cannot be imagined. From equation (2*) and solutions (1), (2'), (4) and (5') at the zero approximation in the fine structure constant $\alpha = \alpha_o = 0$ and finite electron mass $m_e$, it follows $m_\tau$, $m_\mu = \infty$, $m_\nu = 0$ with null electroweak interactions. A very small change of the fine structure constant from $\alpha = 0$ to $\alpha = \alpha_{Data}$ would generate a infinitely large decrease of the muon and tauon masses to their data values together with a very small increase of the neutrino mass scale from zero to $(m_\nu)_{QD} > 0$. Accordingly, without the muon and tauon there would be no bound states of the electron[21] - an important inference from the unique connection of the new universal constant $\alpha_o$ with the CL mass ratios and DMD-hierarchy R, on the one hand, and the electroweak constants $\alpha$ and $\alpha_W$, on the other hand.

*6)* *QD-neutrino mass scale*.
Three estimations of the QD-neutrino mass scale are obtained: 1) from neutrino oscillation data, (32), $m_\nu \cong (7\Delta m^2_{atm})^{1/2} \cong (0.12 \div 0.15)$eV, 2) from neutrino oscillation data (4), see footnote[5], $m_\nu \cong (0.11 \div 0.18)$eV, 3) from drastic lepton mass-ratio hierarchy, (25), (26), $m_\nu \cong (\alpha_o^3 \div \alpha^3)m_e/2^{1/2} \cong (0.11 - 0.14)$eV. They are compatible with the data restrictions on absolute neutrino mass [5-7, 17]. The fairly good agreement between these three entirely independent estimations of the QD-neutrino mass scale is a reassuring quantitative result from the two guiding ideas and CL experimental mass data.

*7)* *Neutrino oscillation mass-squared differences (35) and (36):* $\Delta m^2_{atm} = (\alpha_o^6 \div \alpha^6)m_e^2/14 \cong (1.8 \div 2.8)$ x $10^{-3}$ eV$^2$ and $\Delta m^2_{sol} \cong (5.9 \div 9.5)$ x $10^{-5}$ eV$^2$. The agreement of these predicted ranges of neutrino mass-squared differences with their experimental values from the neutrino oscillation data (32) is a special quantitative test, though still indirect, of the necessary here QD-neutrino type.

---

[21] For illustration only, but see Appendix B for a possible cosmological meaning.

### Appendix A: Dirac-Majorana DMD-opposites

If QD-neutrinos are Dirac particles, the neutrino-CL DMD-oppositeness has no apparent cause. If QD-neutrinos are Majorana particles, the idea of neutrino-CL DMD-oppositeness in the light of known data suggests a Dirac-Majorana DMD-oppositeness condition for all elementary particles in low energy flavor phenomenology including quarks. The final decision is, of course, up to new coming data.

If Dirac-Majorana DMD-oppositeness is a general rule in flavor physics, hypothetical heavy Majorana neutrinos in the seesaw mechanism should also be of QD-type if they really are the right counterparts of the known light left Majorana neutrinos in the Dirac mass terms (comp. [16]).

### Appendix B: Remark on the anthropic selection

The numerical value of the fine structure constant $\alpha$ at zero momentum transfer is shown above to be determined by the exponential $\alpha_o \equiv e^{-5}$ of the integer 5, and this fact may in a sense be the origin of the emphasized in the literature anthropic value $\alpha$ with no use of the anthropic principle.

But if the question 'why is the special value $\alpha_o$ (particular integer 5) singled out in our universe' is a valid one at all, it should have an answer by the anthropic principle: the anthropic principle means here that the *necessary existence of our universe is a kind of spontaneous violation of the basic discrete cosmological symmetry of natural numbers* n = 1,2,3,4,5… The big bang starts with $\alpha = \alpha_o$. The value of the fine structure constant varies with cosmological time in three stages, the second stage is the solution (17) of the Eq.(16) probably related to cosmological inflation and the third one is the contemporary $\alpha$-value as the solution (22*) of the Eq.(21*). Life would be impossible, at least in known forms, in all other imaginable[22] universes defined by the natural numbers **$n \neq 5$**, the universal constant $\alpha_o$ analogues $\alpha_o^{(n)} \equiv e^{-n}$ and fine structure constant $\alpha$ analogues $\alpha^{(n)} \cong \alpha_o^{(n)}$. The spontaneous symmetry violation singles out one unique favorable to life universe, *our* universe with **$n = 5$**, and a new dimensionless fundamental constant $\alpha_o = \alpha_o^{(n)}|_{n=5} = \exp(-5)$ - the source of the electroweak interaction constants $\alpha$, $\alpha_W$ and lepton mass and deviation-from-mass-degeneracy hierarchies, see Conclusions. This uniqueness is in contrast to the infinite numbers of experimentally admissible values of the constant $\alpha$ in the small range $(\Delta\alpha)_{anthrop}$ if the anthropic selection of the value $\alpha$ is made from a background of continuous numbers. The reasoning above implies that there exist a precise connection between the constant $\alpha_o \equiv \exp(-5)$ and the fine structure constant $\alpha(Q^2=0)$. Equation (21*) above is just the needed connection.

---

[22] The term "imaginable" may be thought here in the sense of e.g. the bottom-up approach in fundamental physics: first the equations are found, and then they are derived from the minimal action principle by selections from an imaginable broader system of possible values.

The anthropic principle is sometimes regarded strange to traditional physics; the $\alpha_o$-idea describes this principle in terms of a new fundamental physical constant, and so unites electroweak theory (SM,…) with the idea of anthropic selection of free parameters in traditional physics way.

## Appendix C: Comment on quark mass ratios

Unlike the charged leptons, bare values of the quark masses and mass ratios should be substantially impacted by the renormalization effects of strong interactions. But some essential information may be preserved. Here, an oversimplified attempt is made to address the quark mass ratios by the same two numerical values ($\sqrt{2}/\alpha_o$) and ($\sqrt{2}/\sqrt{\alpha_o}$) that appear in case of the charged leptons (1). The following 2x2 mass-ratio matrix is considered to describe a possible analogy between the bare mass ratios of the Dirac elementary particles, quark and CL, noted in Appendix A:

$$\sqrt{2} \quad \times \quad \begin{vmatrix} 1/\alpha_o & 1/\alpha_o \\ 1/\sqrt{\alpha_o} & 1/\sqrt{\alpha_o} \end{vmatrix} \tag{C1}$$

The numbers in the columns are CL mass ratios ($m_\mu/m_e$) and ($m_\tau/m_\mu$), the numbers in the upper row are the up-quark mass ratios,

$$(m_t/m_c) \cong (m_c/m_u) \cong \sqrt{2}/\alpha_o \cong 210, \tag{C2}$$

while the numbers in the lower row are the down-quark mass ratios

$$(m_b/m_s) \cong (m_s/m_d) \cong \sqrt{(2/\alpha_o)} \cong 17. \tag{C3}$$

By the analogy (C1), the up- and down-quark mass spectra are geometric ones, e.g. [14], in contrast to the case of CL mass spectrum. Nevertheless, the quark mass ratios from (C1) can be represented in the same form (1),(1') as the CL ones, but with small exponents $\chi_q \ll 1$ so that the large values of the quark mass ratios, unlike that of the CL, come not from the exponents, but from large pre-exponential coefficients $\xi_q$:

$$X_{q1} \cong \xi_q \exp\chi_q, \; X_{q2} \cong \xi_q \exp(\chi_q/2), \; \chi_q \ll 1,$$

$$\xi_{up} \cong \sqrt{2/\alpha_o}, \; \xi_{down} \cong \sqrt{(2/\alpha_o)}. \tag{C4}$$

At the considered approximation, the small quark exponents $\chi_q$ have trivial meaning.

Excluding gravity, the considered three groups of elementary particles (neutrinos, charged leptons and quarks) participate respectively in three fundamental interactions (weak, electromagnetic and strong), as the dominant ones. As pointed out above, comp. (43), the two DMD-hierarchies of the neutrinos $r$ and CL R are links between the neutrino oscillation and lepton mass data, on the one hand, and the two basic weak and electromagnetic dimensionless coupling constants,

$$r \cong \alpha_{oW} \cong \alpha_W, \; R \cong \alpha_o \cong \alpha, \tag{C5}$$

on the other hand. With (C1), and notations for mass-ratio-squared hierarchies of the up- and down-quarks,

$$R_u \equiv (m_t/m_c)^2/(m_c/m_u)^2 \;\;, \;\; R_d \equiv (m_b/m_s)^2/(m_s/m_d)^2, \tag{C6}$$

an extension of the lepton succession (C5) is given by

$$r \sim \alpha_W, \; R \sim \alpha, \; R_u \sim \alpha_s \sim 1, \; R_d \sim \alpha_s \sim 1, \tag{C7}$$

compare footnote[17]. The unities in (C7) for the quark mass ratio hierarchies indicate the geometrical structure of the quark mass spectra; they should be related to the strong interactions. Note that in contrast to the CL there are two types of quark mass-ratio-squared hierarchies from the matrix (C1): 'horizontal' hierarchies included in (C7) and related to the geometrical spectra of the up- and down-quarks, and 'vertical' hierarchies between up- and down-quarks mass ratios; the latter are identical to the mass-ratio hierarchy of the CL, equal $\alpha_\circ$.

With the mass ratios in (C2) and (C3), the data value [8] of the t-quark mass $m_t \sim 175$ GeV leads to current up-quark masses $m_c \sim 0.8$ GeV and $m_u \sim 4$ MeV. The data value of the b-quark mass $m_b \sim 4.5$ GeV leads to current down-quark masses $m_s \sim 260$ MeV and $m_d \sim 15$ MeV. These estimations disagree with the data values [8] by about a factor of $\leq 2$.